**Trends in Open Access Academic Outputs of State Agricultural Universities in India: Patterns from OpenAlex**

Abhijit Roy[a], Akhandanand Shukla[b] and Aditya Tripathi[c]

[a] Senior Research Fellow, Department of Library & Information Science, Central University of Tamil Nadu,
Thiruvarur, Pin 610005, India. E-mail: royabhijit75@gmail.com

[b] Associate Professor, Department of Library & Information Science, Central University of Tamil Nadu, Thiruvarur, Pin 610005, India. E-mail: akhandanandshukla@cutn.ac.in

[c] Professor, Department of Library & Information Science, Banaras Hindu University, Varanasi, Pin 221005, India.
E-mail: aditya.tripath@gmail.com

## 1. Introduction and review of literature

Agriculture forms the foundation of India's civilization, culture and heritage. About 60% of the people in India live in rural areas and derive their livelihood from agriculture and allied sources (Gulati & Juneja, 2022; Pathak et al., 2022). After independence, different premiere agriculture revolutions (Round Revolution- Potato; Green revolution-rice and wheat; Grey Revolution- Fertilizers/wool production; Pink Revolution- Prawn or onion production; White Revolution- Milk production; Blue Revolution- Fish production; Red Revolution- Meat or tomato production; Yellow revolution- Oilseed production; Brown revolution- Leather/cocoa production; Golden fibre revolution- Jute production; Golden revolution- Fruits/Honey/Horticulture production; Silver revolution- Egg/Poultry production; Silver fibre revolution- Cotton; and Evergreen revolution for overall development of agriculture) were transformed India from a food importer to food exporter country (Goswami et al., 2017). However, it would be a major challenge for Indian agriculture to feed this large population, climate change, degradation of natural resources, and Insect infestations. Indian Council of Agriculture Research (ICAR) is an autonomous organisation under the Department of Agriculture Research and Education (DARE), Ministry of Agriculture and Farmers Welfare, Government of India. It was established on 16 July 1929. It has sixty-six research institutions, eleven Agricultural Technology Application Research Institutes, twelve Project Directorates, six National Bureaux, and fifteen National Research Centres. Moreover, three types of universities- three central universities, four deemed universities, and sixty-three state universities- work together overall in India. This is the largest agriculture system in the world (https://icar.org.in/about-us).

Scientometric and bibliometric in both fields measure the scholarly publishing scientific growth and impacts. The performance of countries, institutions and scholars in a subject, field or domain. It also finds out scholarly communication tendencies and information-seeking behaviours of scholars. (Balasubramanian & Ravanan, 2011; Nayak & Bankapur, 2017; Sagar et al., 2013) examined the growth of global agriculture research based on different databases. They adopted the Scientometric approaches and visualised the productive country, journals, publishers, research fields, and major keywords. In addition, the authors visualize the agriculture research performance of India, Bangladesh, Zambia, Nigeria, Pakistan, Tanzania and Malaysia (Chisenga & Simumba, 2009; Das et al., 2019; Garg et al., 2011; Kasa et al., 2014; Nasir et al., 2019; Pouris, 1989;

Vellaichamy, 2016). However, some bibliometric studies explore the ICAR-affiliated institutions, Universities and research centres' scientific outputs, country collaborations, journals, authors, trending topics and keywords (Behera et al., 2022; Garg et al., 2006; Kadam & Bhusawar, 2021; Nidhisha & Sarangapani, 2021; Ramanan S. et al., 2021; Sankar & Prema, 2022; Sarkhel & Raychoudhury, 2010; Sharma, 2009; Tekale et al., 2017; Tripathi et al., 2013).
In the last two decades, with the advance of the internet and ICT applications, there has been a major shift in scholarly communication. Scholars submitted papers in OA routes and got advantages for scholarly impacts (Uma & Kumar, 2015). The five major OA routes are visible in recent scholarly communications. Green OAs published a copy in the Archive, personal website and blog (Gargouri et al., 2012); Diamond OA published OA journals (Fuchs & Sandoval, 2013). Gold OA was published in an OA journal with Article processing charges (Archambault et al., 2014; Piwowar et al., 2018). Hybrid OA published a closed-access journal with a legal licence. It also charges APC from Authors (Björk, 2012). The last one, Bronze OA, is free for readers only on the publishers' website without any licensing information (Laakso & Björk, 2013). Out of those, other types of publications are closed access.
OpenAlex is a global open-access scholarly catalogue made by OurResearch, non-profit organization (Priem et al., 2022). It describes seven entities, i.e., Works, authors, sources, institutions, topics, publishers and funders. It is a big, open and easy-to-assess database. Scholars would analyze and access world research in a simple way. It provides free REST API and data snapshot services (Hazarika et al., 2024). This paper highlighted the open-access academic outputs of Indian state agriculture universities. However, the government of India adopted the OA policies. In 2009, the National Knowledge Commission recommended that all educational resources be open. The Council of Scientific & Industrial Research (CSIR) constituted the comments in 2011 for the implementation of an open access policy for CSIR labs. In the same year, UGC and INFLIBNET centre launched the Sodhganga: a repository for theses of Indian universities. The Indian Council of Agriculture Research (ICAR) adopted the OA policy and built up the institutional repositories. It also provides funding support for innovative research works. Likewise, the Department of Biotechnology (DBT) and the Department of Science and Technology (DST) jointly provide funding support for OA publishing. As a result, previous studies revealed that India is a responsible country for sharing OA publications globally (Robinson-Garcia et al., 2020). (Nazim, 2021) observed that in Indian institutions 23 percent of publications share in OA routs. Likewise, authors focus on the Open Access friendliness of Indian IITs, Central Universities, State Universities, NITs and Non-profit Organisations (Hazarika et al., 2024; Mukhopadhyay, 2022; Roy & Mukhopadhyay, 2022b, 2022c, 2022a).
In view of the above situations, the present study examined the open-access academic outputs of Indian state agriculture universities. The data carpentry tool OpenRefine is employed for data cleaning and data analysis. The ArcGIS 10.8 version is used to create the maps of OA publishing. The research fulfils the following objectives

1. To reveal the state of OA Indian state Agriculture Universities in the last ten years (2014-2023)
2. To examine the top productive sources.
3. To observe the country, institution, authors' collaboration and scholarly impact.
4. Focus on funding received and Article processing charge using tendency and scholarly impacts.
5. To find out the top useable keywords and
6. To figure out the OA publications related to domain and trending topics

7. Examine the publication's locations and impact.

The experiment fills the gap in the existing research by providing a comprehensive overview of the open-access publishing landscape of Indian state agriculture universities. Authors, institutions and policymakers would review the current scenario and make decisions for future improvement. The remaining sections of the articles are organized as follows: Section 2: research methodology using data carpentry approaches, section 3: data analysis and data interpretation, section 4 discusses and findings and Section 5: conclusion of the study.

## 2. Methodology

*2.1 Selection of Institutions*

The investigations reveal the OA academic outputs of Indian state agriculture universities based on the OpenAlex database. A total of 63 state agriculture universities are affiliated with ICAR (https://icar.org.in/state-agricultural-universities).

*2.2 Data collection and cleaning*

The research includes, The Research Organization Registry (ROR ID) for data collection to OpenAlex. ROR ID is a global community-led registry of open persistent identifiers for research organizations. It makes it easy to identify a particular institution and its scholarly outputs. Among 63 universities, seven (Bihar Animal Sciences University, Patna; Junagarh Agricultural University, Junagarh; Haryana State University of Horticultural Sciences, Karnal; Rajmata Vijayaraje Scindia Krishi Vishwa Vidyalaya, Gwalior; Agriculture University, Kota; Agriculture University, Jodhpur and Banda University of Agricultural and Technology, Banda) universities are not listed in ROR. OpenAlex also does not provide these institutions with scholarly outputs. So, all six institutions were excluded from further study.

In 56 universities, academic outputs were filtered within the last ten years (2014-2023) and downloaded in CSV files—the data carpentry tool OpenRefine employed for data cleaning and marge in data sets (Hazarika et al., 2024; Roy & Shukla, 2024). A total of 97881 records were downloaded from OpenAlex (Priem et al., 2022). It was found that 345 records were not relevant. They were so excluded from further study. Figure 1 displays the details of the methodology in the Flowchart.

When the experiments were checked the data sets, it was found that 67.55% (65889 out of 97536) of academic outputs were distributed in five OA routes i.e., Green, Gold, Dimond, Hybrid and Bronze. Moreover, 32.45% (n=31647) were distributed in Closed Access.

However, the following REST APIs are used for Authors, country collaborations, keywords and journal details-

1)
https://api.openalex.org/works?page=1&filter=authorships.institutions.lineage:i12419238|i3131536999|i3130939756|i35839044|i2800696116|i20632288|i2800089517|i271267010|i34750443|i152185767|i3133099055|i24927338|i233492666|i377099284|i211055903|i205478545|i1306687154|i235110982|i199921877|i244504473|i61553790|i139046659|i102458738|i4210150675|i54021868|i221342038|i335685885|i4210156499|i290859477|i2799533726|i3130477807|i120386524|i167494164|i223781097|i13567498|i223471776|i899646|i253568910|i2801834944|i4210128267|i4210128752|i134900695|i82452031|i304343950|i3132430671|i4210098074|i4210092736|i252758333|i4210095517|i4210143625|i4210097692|i4210098184|i4210141180|i109963156|i75444546|i4210116187,publication_year:2014-2023,open_access.is_oa:true&apc_sum=false&cited_by_count_sum=false

2) "https://api.openalex.org/sources/" + value

*2.3 Data visualizations*

After data analysis, the research used the ArcGIS 10.8 version tool and MS Excel for data visualization.

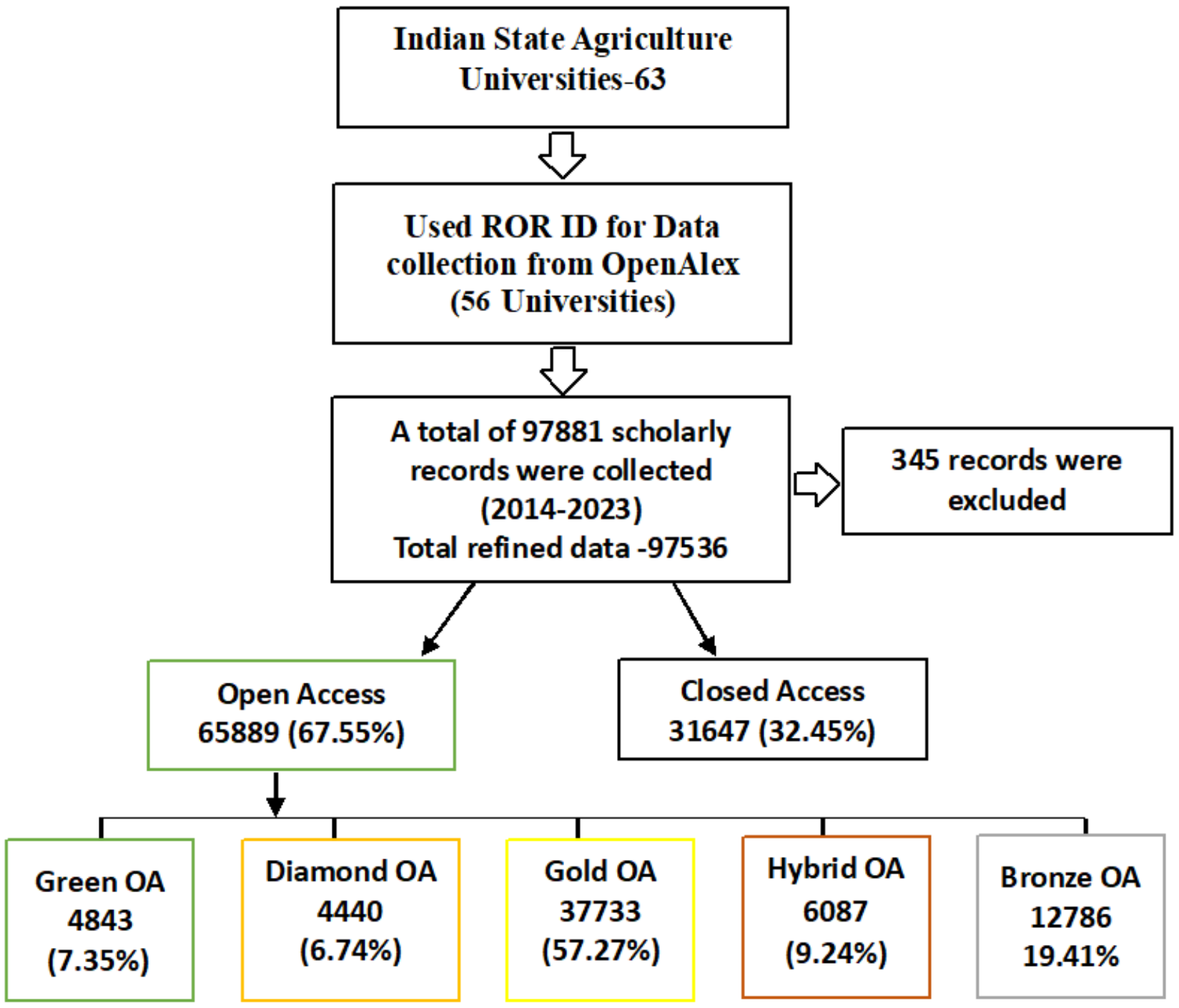

Figure 1: Flowchart of research workflow

## 3. Data analysis

*3.1 Growth of OA publication*

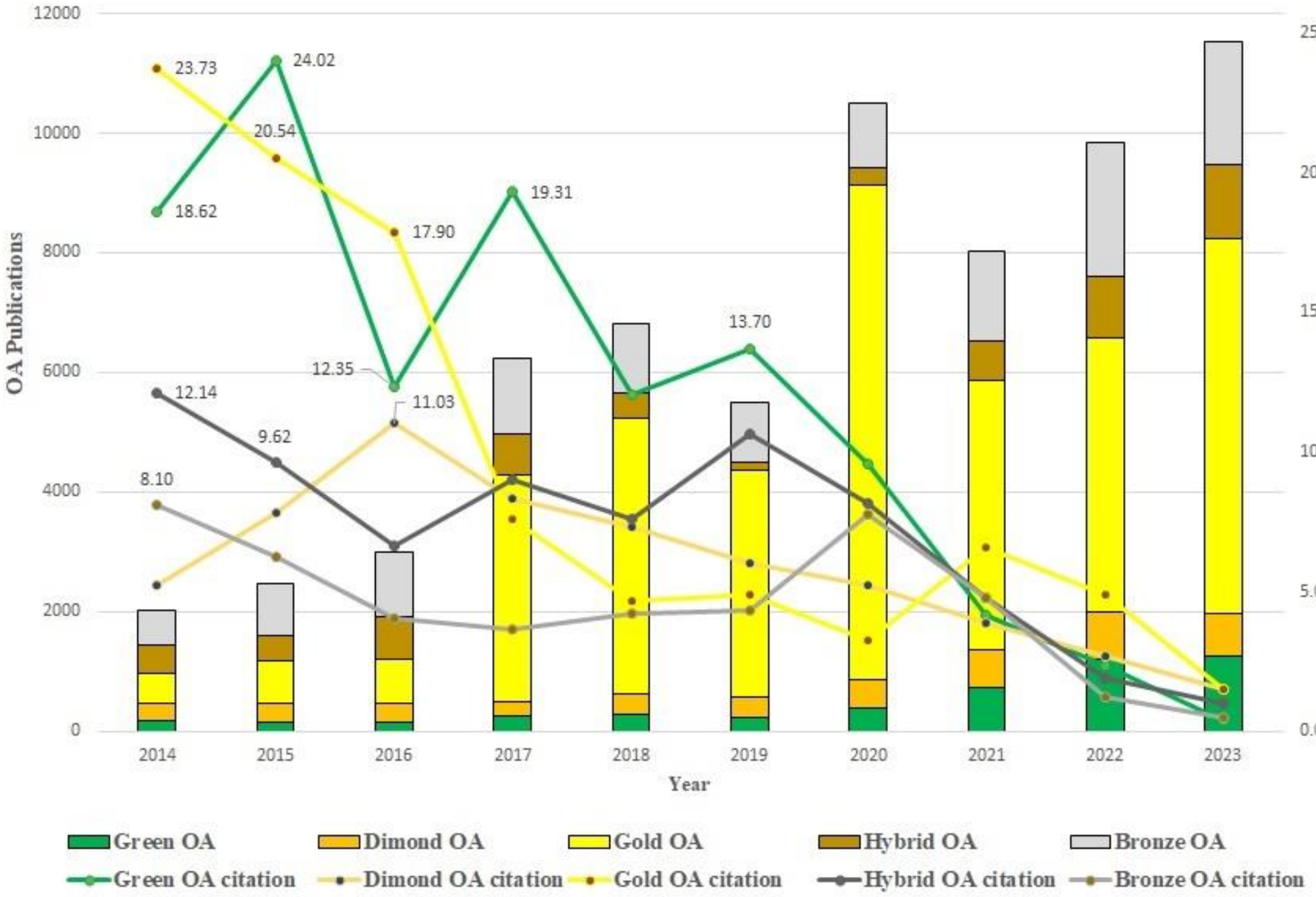

Figure 2: OA publications growth in 2014-2023

The experiments highlighted the yearly OA publication's growth in Figure 2. The numbers of OA publications and average citations were presented. In the last ten years (2014-2023), a total of 65,889 OA academic research outputs were distributed by fifty-six state universities in five OA routes. Gold OA is the most productive OA, with 57.27% of records shared in this route and the average citation per paper was 5.21. followed by Bronze OA 19.41% with ACP 3.70, Hybrid OA 9.24% with ACP 5.64, Green OA 7.35% with ACP 6.32 and Dimond OA 6.74% with ACP 4.93. The publication growth increased continuously from 2017 in 2014 to 11535 in 2023. In the year 2023, it shared the highest 11535 OA academic outputs. In 2020, it was published (10513) records as the second-highest publication. The figure reveals that in 2015, the highest average citations per paper (24.02) came from green access papers. Green OA and gold OA papers citations are ahead of the other three routes. The authors' shelf archiving policy gives more citation advantages in Green OAs. In the last ten years, only Green OA documents have an average of citations per document above 11.65.

*3.2 OA Status of State Agriculture Universities*

Table 1 tabulated the 56 State Agriculture universities' OA outputs, OA colours and time citations according to their OA publishing. It listed 27 universities' names that OA shares above a thousand. Tamil Nadu Agricultural University is the highest OAs productive institution. In the last ten years, it published 5709 (8.66% of total OA) academic papers (Green OA- 322, Gold OA-3094, Diamond-383, Hybrid-802 and Bronze OA-1108) with 22392 (6.77% of total citations) citations. However, Punjab Agricultural University (OA-3478; TC-31127), Chaudhary Charan Singh Haryana Agricultural University (OA-2967; TC-13939), G.B. Pant University of Agriculture & Technology (OA-2938; TC-31116) and Sher-e-Kashmir University of Agricultural Science & Technology, Srinagar (OA-2503; TC-22240) each were published above two thousand OA papers and took place in the top five ranks. In OA Categories, Tamil Nadu Agricultural University distributed the highest numbers of Gold OA- 3094, Diamond OA – 383, Hybrid OA-802 and Bronze OA- 1108. The highest numbers of Green OA- 361 and Time citations-31127 came from Punjab Agricultural University. However, U.P. Pt. Deen Dayal Upadhyaya Pashu Chikitsa VigyanVishwaVidhyalaya Evem Go Anusandhan Sansthan, Mathura published 456-OA papers and received TC-12408. Its average citation per paper is 27.21.

Table 1: OA status of State Agriculture Universities based on OpenAlex in 2014-2023

| Institution Name | No. of OAs | OA Types (%) | | | | | TC | ACP |
|---|---|---|---|---|---|---|---|---|
| | | Green | Gold | Diamond | Hybrid | Bronze | | |
| 1) Tamil Nadu Agricultural University | 5709 | 5.64 | 54.20 | 6.71 | 14.05 | 19.41 | 22392 | 3.92 |
| 2) Punjab Agricultural University | 3478 | 10.38 | 39.76 | 9.75 | 16.65 | 23.46 | 31127 | 8.95 |
| 3) Chaudhary Charan Singh Haryana Agricultural University, Hisar | 2967 | 5.86 | 54.94 | 8.70 | 11.46 | 19.04 | 13939 | 4.70 |
| 4) G.B. Pant University of Agriculture & Technology | 2938 | 8.71 | 51.46 | 8.61 | 12.70 | 18.52 | 31116 | 10.59 |
| 5) Sher-e-Kashmir University of Agricultural Science & Technology, Srinagar | 2503 | 10.67 | 57.89 | 6.63 | 8.79 | 16.02 | 22240 | 8.89 |
| 6) Assam Agricultural University | 1948 | 7.34 | 55.80 | 6.47 | 6.26 | 24.13 | 6714 | 3.45 |
| 7) Orissa University of Agricultural & Technology, Bhubaneswar | 1944 | 6.84 | 59.62 | 5.92 | 8.85 | 18.78 | 8785 | 4.52 |
| 8) Professor Jayashankar Telangana State Agricultural University, Hyderabad | 1755 | 3.42 | 71.85 | 3.30 | 4.56 | 16.87 | 4892 | 2.79 |
| 9) University of Agricultural Sciences, Dharwad | 1753 | 4.79 | 58.13 | 3.59 | 5.53 | 27.95 | 7407 | 4.23 |
| 10) Jawaharlal Nehru Krishi Viswa Vidyalaya, Jabalpur | 1673 | 3.65 | 68.32 | 4.06 | 5.98 | 17.99 | 4464 | 2.67 |
| 11) University of Agricultural Sciences, Bangalore | 1583 | 6.95 | 61.78 | 3.98 | 8.02 | 19.27 | 10284 | 6.50 |
| 12) Bidhan Chandra Krishi Viswa Vidhyalaya, Mohanpur | 1572 | 7.89 | 52.74 | 5.22 | 14.19 | 19.97 | 10286 | 6.54 |
| 13) Maharana Pratap University of Agriculture & Technology, Udaipur | 1557 | 4.88 | 55.68 | 5.46 | 7.06 | 26.91 | 7585 | 4.87 |
| 14) Tamil Nadu Veterinary & Animal Sciences University, Chennai | 1544 | 10.30 | 57.45 | 8.74 | 8.61 | 14.90 | 8527 | 5.52 |
| 15) Acharya NG Ranga Agricultural University, Guntur | 1508 | 7.43 | 56.23 | 4.24 | 7.76 | 24.34 | 4169 | 2.76 |
| 16) Kerala Veterinary and Animal Sciences University, Pookode | 1466 | 10.16 | 55.87 | 6.82 | 9.00 | 18.14 | 6347 | 4.33 |
| 17) Anand Agricultural University, Anand | 1457 | 6.66 | 58.13 | 4.80 | 12.90 | 17.50 | 7583 | 5.20 |
| 18) Dr. Yaswant Singh Parmar University of Horticulture & Forestry, Solan | 1418 | 5.71 | 55.71 | 7.76 | 10.30 | 20.52 | 7201 | 5.08 |
| 19) Guru Angad Dev Veterinary and Animal Sciences University, Ludhiana | 1375 | 12.58 | 46.33 | 16.36 | 8.44 | 16.29 | 9607 | 6.99 |
| 20) Swami Keshwanand Rajasthan Agricultural University, Bikaner | 1291 | 3.41 | 57.47 | 7.75 | 6.43 | 24.94 | 3832 | 2.97 |
| 21) Kerala Agricultural University, Thrissur | 1241 | 9.35 | 56.00 | 7.09 | 8.54 | 19.02 | 4261 | 3.43 |
| 22) Sher-e-Kashmir University of Agricultural Science & Technology, Jammu | 1230 | 6.91 | 59.43 | 10.33 | 7.15 | 16.18 | 9776 | 7.95 |
| 23) Navsari Agricultural University, Navsari | 1188 | 3.37 | 57.74 | 6.14 | 10.77 | 21.97 | 3729 | 3.14 |
| 24) Indira Gandhi Krishi Viswa Vidhyalaya, Raipur | 1141 | 2.72 | 72.13 | 2.89 | 5.43 | 16.83 | 3376 | 2.96 |
| 25) University of Agricultural Sciences, Raichur | 1131 | 5.13 | 70.03 | 1.95 | 9.20 | 13.70 | 3249 | 2.87 |
| 26) Vasantrao Naik Marathwada Krishi Vidyapeeth, Parbhani | 1098 | 3.73 | 61.66 | 1.91 | 3.83 | 28.87 | 2067 | 1.88 |
| 27) Ch. Sarwan Kumar Himachal Pradesh Krishi Viswavidyalaya, Palampur | 1000 | 11.20 | 49.70 | 9.70 | 10.50 | 18.90 | 4459 | 4.46 |
| Three institutions' publications 900-999 OAs | 2846 | 3.44 | 68.55 | 3.72 | 8.12 | 16.16 | 8120 | 2.85 |
| Two institutions publications -800-899 OAs | 1714 | 0.04 | 64.70 | 5.13 | 5.02 | 20.95 | 4839 | 2.82 |
| Three institutions publications -700-799 OAs | 2234 | 0.10 | 54.16 | 9.67 | 6.45 | 19.70 | 11493 | 5.14 |
| Four institutions publications -600-699 OAs | 2627 | 0.08 | 63.30 | 6.47 | 6.13 | 16.41 | 9085 | 3.46 |
| Four institutions publications -500-599 OAs | 2252 | 0.07 | 61.63 | 6.57 | 6.75 | 17.94 | 7506 | 3.33 |
| Four institutions publications -400-499 OAs | 1875 | 0.13 | 56.59 | 9.17 | 7.04 | 13.97 | 17895 | 9.54 |
| Seven institutions publications -300-399 OAs | 2520 | 0.14 | 48.93 | 7.94 | 10.44 | 18.65 | 9641 | 3.83 |
| Two institutions publications -158-299 OAs | 353 | 0.05 | 66.86 | 4.53 | 6.52 | 17.56 | 2801 | 7.93 |
| Sub-Total of 56 Agriculture Universities OA publications | 65889 | 7.35 | 57.27 | 6.74 | 9.24 | 19.41 | 330794 | 5.02 |

*TC-Time Citation, ACP-Average citations per paper*

### *3.3 Growth of OA in the last ten years in India*

Figure 3 highlights yearly OA publications and growth in the state and union territories. The colour yellow to deep brown indicates low to high numbers of OA publishing in the state and territory. In all 56 state Universities were established in 20 states and one union territory. In the last ten years, the highest 7624 (11.57% of total OA) numbers of OA come from Tamil Nadu. Followed by Karnataka -6229 (9.45%), Punjab -4853 (7.37%), 5.87%-5.23% of OAs come from Rajasthan,

Jammu and Kashmir, Haryana, Gujarat, Maharashtra, 4.79%-4.15% of OAs published by Uttar Pradesh, Uttarakhand, Kerala, West Bengal, Andhra Pradesh; 3.77%-0.56% of OAs distributed from Telangana, Himachal Pradesh, Madhya Pradesh, Assam, Odisha, Chhattisgarh, Bihar and Jharkhand. However, if we observe the yearly growth, the highest 226 of OA records were published in Punjab in 2014 (Figure 3a). In the years 2017, 2018 and 2019 the highest numbers (730, 831, 626) OAs were shared by Karnataka Agriculture universities (Figure 3d,e & f). In the remaining six years (2015, 2016, 2020, 2021, 2022 and 2023) the highest numbers of OAs were published in Tamil Nadu (Figure 3c, g, h, i & j).

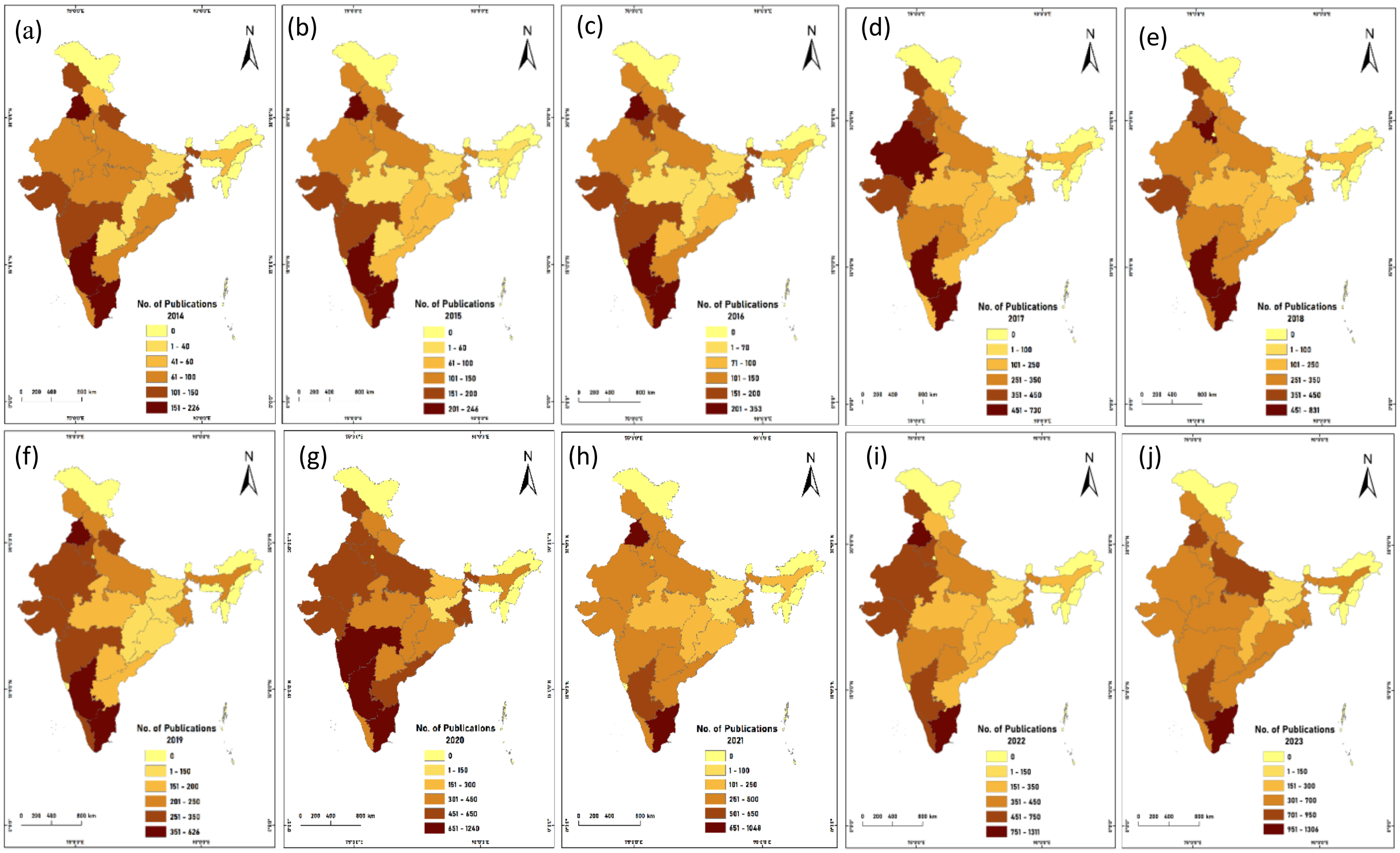

*Figure 3: OA growth at the state level in the year 2014-2023*

### *3.4 Productive Sources*

#### *3.4.1 Domestic sources*

Of the 65889 OAs, academic outputs were distributed by six primary sources, i.e., journal articles 61508 (93.35%), Repository-4.55%, Others-1.53%, ebook platform - 0.34% book series -0.13% and Conference- 0.09%. Table 2 shows that of the 61508 journal articles, 78.34% were published by 864 domestic journals. More than five hundred OA publishing top 14 journals names, OA numbers, Average citation, journal h-index, i-10 index and journal quartile were highlighted. The highest number (13458) of OA was published by the International Journal of Current Microbiology and Applied Sciences. Excellent Publisher's agencies published it. Among the 14 journals, the highest (13.61) number of Average citations per paper comes from the Veterinary World journals. It received the highest 57 h-index, 1252 i-10 index and Quartile 2. It is published by Veterinary World. Of the 14 journals, 50% (n=7) were not listed in the Scopus. However, Scopus listed seven journals; two (Journal of Applied and Natural Science Indian & Journal of Agricultural Sciences) are in quartile 4 groups, 4 are in quartile 3 groups and only one is listed in Quartile 2.

Table 2: journals affiliated with India

| Rank | Journal Name | OA Status | ACP | h-index | i-10 index | Q |
|---|---|---|---|---|---|---|
| 1 | International journal of current microbiology and applied sciences | 13458 | 2.22 | 41 | 791 | NL |
| 2 | International journal of chemical studies | 2675 | 0.76 | 25 | 85 | NL |
| 3 | International Journal of Environment and Climate Change | 2581 | 0.40 | 10 | 11 | NL |
| 4 | International journal of plant and soil science | 1929 | 0.47 | 20 | 86 | NL |
| 5 | Journal of applied and natural science | 1426 | 3.19 | 29 | 187 | 4 |
| 6 | Indian journal of agricultural sciences | 1139 | 1.57 | 30 | 395 | 4 |
| 7 | Indian journal of animal sciences | 1115 | 1.30 | 23 | 190 | 3 |
| 8 | Current journal of applied science and technology | 1076 | 1.04 | 17 | 44 | NL |
| 9 | Asian journal of agricultural extension, economics and sociology | 920 | 0.32 | 16 | 43 | NL |
| 10 | Indian Journal of Animal Research | 804 | 1.03 | 17 | 67 | 3 |
| 11 | Legume Research | 751 | 1.58 | 21 | 170 | 3 |
| 12 | Pharma innovation | 703 | 0.63 | 10 | 9 | NL |
| 13 | Electronic journal of plant breeding | 546 | 2.06 | 20 | 74 | 3 |
| 13 | Veterinary world | 546 | 13.61 | 57 | 1252 | 2 |
| | 9 journals publishing 300-499 papers each | 3677 | | | | |
| | 15 journals publishing 200-299 papers each | 3727 | | | | |
| | 28 journals publishing 100-199 papers each | 4110 | | | | |
| | 39 journals publishing 50-99 papers each | 2616 | | | | |
| | 34 journals publishing 30-49 papers each | 1303 | | | | |
| | 26 journals publishing 20-29 papers each | 620 | | | | |
| | 71 journals publishing 10-19 papers each | 930 | | | | |
| | 92 journals publishing 5-9 papers each | 594 | | | | |
| | 47 journals publishing 4 papers each | 188 | | | | |
| | 79 journals publishing 3 papers each | 237 | | | | |
| | 103 journals publishing 2 papers each | 206 | | | | |
| | 307 journals publishing 1 paper each | 307 | | | | |
| | Subtotal | 48184 | | | | |

*3.4.2 Foreign journals*

Table 3 shows the names of fourteen foreign journals that were published in over a hundred articles. Of the 61508 journal articles, 21.66% (n=13324) were published by 2776 journals from 83 countries. The top three countries are the United States -611 journals, the United kingdom-451 journals, and the Netherlands- 271 journals. The Frontiers in Plant Science journal is the top productive journal. It published 396 articles and ACP 29.32. Of the fourteen journals, 6 are affiliated with Switzerland, two are affiliated with the United Kingdom and six are affiliated with six different countries. Among the 14 journals, the highest 48.36 ACP comes from a Switzerland-affiliated journal the Frontiers in Microbiology. It is a quartile-1 listed journal. Of the listed journals, nine are listed in Quartile 1, two are listed in Quartile 2, one journal is listed in Quartile 3, another is in Quartile 4 and only one is not listed in Scopus.

*Table 3:* Journals affiliated with foreign

| Rank | Journal Name & Affiliated Country | OA Status | ACP | h-index | i-10 index | Q |
|---|---|---|---|---|---|---|
| 1 | Frontiers in plant science-Switzerland | 396 | 29.32 | 249 | 15237 | 1 |
| 2 | Scientific reports-United Kingdom | 337 | 24.93 | 362 | 118653 | 1 |
| 3 | PloS one-United States | 280 | 27.47 | 513 | 202429 | 1 |
| 4 | 3 biotech-Switzerland | 223 | 28.20 | 89 | 1803 | 2 |
| 5 | Sustainability- Switzerland | 221 | 21.60 | 211 | 31435 | 1 |
| 6 | Agronomy- Switzerland | 217 | 18.40 | 112 | 4212 | 1 |
| 7 | Journal of food processing and preservation-UK | 179 | 9.21 | 78 | 2981 | 2 |
| 8 | Bangladesh journal of botany- Bangladesh | 176 | 2.96 | 24 | 129 | 4 |
| 9 | African journal of agricultural research-Nigeria | 133 | 9.18 | 62 | 1451 | NL |
| 10 | Frontiers in microbiology- Switzerland | 132 | 48.36 | 272 | 19221 | 1 |
| 11 | Plants- Switzerland | 125 | 14.33 | 111 | 4565 | 1 |
| 12 | Egyptian Journal of Biological Pest Control-Egypt | 120 | 8.98 | 32 | 222 | 1 |
| 13 | Buffalo Bulletin- Thailand | 104 | 0.16 | 13 | 22 | 3 |
| 13 | Heliyon- Netherlands | 104 | 10.49 | 113 | 6228 | 1 |
| | 17 journals publishing 50-99 papers each | 1087 | | | | |
| | 34 journals publishing 30-49 papers each | 1255 | | | | |
| | 176 journals publishing 10-29 papers each | 2927 | | | | |
| | 261 journals publishing 5-9 papers each | 1693 | | | | |
| | 134 journals publishing 4 papers each | 536 | | | | |
| | 230 journals publishing 3 papers each | 690 | | | | |
| | 479 journals publishing 2 papers each | 958 | | | | |
| | 1431 journals publishing 1 paper each | 1431 | | | | |
| | Subtotal | 13324 | | | | |

*3.5 Country collaborations*

Table 4 shows the names of the top twenty collaborative countries with OA colours, time citations, and average citations per paper. Indian State Agriculture Universities collaborated OA papers with 159 countries. The highest 2080 scholarly records (Green -256, Diamond 110, Gold-1107, Hybrid-199 and Bronze-408) were collaborated with the USA and received the highest 33751 citations. Followed by China, Saudi Arabia, the UK, Australia and Czechia, these six countries also collaborated over five hundred papers. In the USA, China, Saudi Arabia, the UK and Australia in five countries collaborative papers received citations of over ten thousand. However, among the top twenty countries, Korea and Australia in both countries collaborative papers average citations per paper over 29. Out of the two countries, the five countries (the UK, Egypt, Germany, Bangladesh and Canada) collaborative papers received ACP over 20.00. Figure 4a presents that 86.23% of OA documents collaborated with a single country. However, Figure 4b highlighted that single-country collaborated papers received 3.56 ACP. But it increased continuously when the numbers of countries increased (2 countries-9.51, 3-18.49, 4-22.56, 5-28.08, 6-29.18, 7-32.29, 8-37.17). it was highly incised when 24 papers were collaborated with nine countries and received ACP 65.00.

*Table 4: collaborative countries*

| SL | Country Name | OA Status | | | | | OAs | Time Citations | ACP |
|---|---|---|---|---|---|---|---|---|---|
| | | Green | Diamond | Gold | Hybrid | Bronze | | | |
| 1 | USA | 256 | 110 | 1107 | 199 | 408 | 2080 | 33751 | 16.23 |
| 2 | China | 84 | 31 | 552 | 83 | 132 | 882 | 14437 | 16.37 |
| 3 | Saudi Arabia | 61 | 14 | 539 | 44 | 95 | 753 | 13436 | 17.84 |
| 4 | UK | 98 | 17 | 354 | 82 | 86 | 637 | 14519 | 22.79 |
| 5 | Australia | 80 | 12 | 309 | 104 | 53 | 558 | 16199 | 29.03 |
| 6 | Czechia | 28 | 20 | 370 | 47 | 77 | 542 | 5593 | 10.32 |
| 7 | Russia | 21 | 11 | 331 | 39 | 70 | 472 | 5871 | 12.44 |
| 8 | Japan | 47 | 11 | 220 | 61 | 88 | 427 | 6781 | 15.88 |
| 9 | Egypt | 31 | 17 | 232 | 22 | 89 | 391 | 8345 | 21.34 |
| 10 | Poland | 12 | 9 | 242 | 41 | 45 | 349 | 4052 | 11.61 |
| 11 | Germany | 40 | 8 | 174 | 64 | 28 | 314 | 7359 | 23.44 |
| 12 | Korea | 24 | 14 | 175 | 49 | 42 | 304 | 9024 | 29.68 |
| 13 | Pakistan | 39 | 7 | 173 | 22 | 60 | 301 | 5897 | 19.59 |
| 14 | Italy | 46 | 10 | 118 | 46 | 77 | 297 | 3246 | 10.93 |
| 15 | Hungary | 7 | 20 | 212 | 11 | 27 | 277 | 2376 | 8.58 |
| 16 | Spain | 55 | 15 | 138 | 35 | 29 | 272 | 5073 | 18.65 |
| 17 | Bangladesh | 16 | 5 | 177 | 28 | 24 | 250 | 6313 | 25.25 |
| 18 | Canada | 36 | 7 | 134 | 26 | 38 | 241 | 4902 | 20.34 |
| 19 | France | 46 | 21 | 113 | 26 | 31 | 237 | 4606 | 19.43 |
| 20 | Sweden | 12 | 7 | 83 | 13 | 63 | 178 | 2802 | 15.74 |
| | Sub Total | 1039 | 366 | 5753 | 1042 | 1562 | 9762 | 174582 | |

### *3.6 Prolific Authors*

A total of 315254 authors worked together on 65889 scholarly works. Table 5 presented the top ten OA productive authors' names, OA types, Time citations, ACP and authors' h-index. Kuldeep Dhama is the top OA productive author from U.P. Pt. Deen Dayal Upadhyaya Pashu Chikitsa VigyanVishwaVidhyalaya Evem Go Anusandhan Sansthan, Mathura. He shared 211 OA papers (Green-22, Diamond-10, Gold-119, Hybrid-15 and Bronze-49) and received 12018 time citations (ACP-55.90 and h-index-81). Of the ten authors, Kuldeep Dhama and Ruchi Tiwari's Time citations are over ten thousand; also, both have average citations of over 55. Figure 4a points out that the authors with the highest 15.90% and 15.88% work together in five and four authorship groups. However, ACP increased when authorship patterns increased. It highlighted nine authorship patterns -10.48, ten authorship patterns 12.22 and the highest 21.14 ACP comes from 10+ authorship patterns (Figure 4b).

*Table 5: Top OA productive authors*

| Authors name | OA status | | | | | Total OAs | Time Citations | ACP | h-index |
|---|---|---|---|---|---|---|---|---|---|
| | *Green* | *Diamond* | *Gold* | *Hybrid* | *Bronze* | | | | |
| Kuldeep Dhama | 22 | 10 | 119 | 15 | 49 | 211 | 12018 | 55.90 | 81 |
| M. Raveendran | 15 | 25 | 93 | 13 | 12 | 158 | 1451 | 9.18 | 27 |
| Ruchi Tiwari | 12 | 5 | 76 | 12 | 34 | 139 | 10406 | 74.86 | 67 |
| Chaitanya G. Joshi | 19 | 2 | 72 | 15 | 5 | 113 | 1438 | 12.73 | 19 |
| Yashpal Singh Malik | 17 | 10 | 51 | 9 | 24 | 111 | 4934 | 44.45 | 45 |
| Rajeev K. Varshney | 30 | 2 | 55 | 12 | 3 | 102 | 3620 | 35.49 | 114 |
| Shabir Hussain Wani | 17 | 1 | 59 | 5 | 13 | 95 | 3301 | 34.75 | 37 |
| V. Geethalakshmi | 4 | 10 | 56 | 15 | 7 | 92 | 145 | 1.58 | 8 |
| N. Senthil | 1 | 9 | 62 | 11 | 7 | 90 | 782 | 8.69 | 22 |
| Z. A. Dar | 4 | -- | 65 | 7 | 12 | 88 | 1139 | 12.94 | 7 |

*3.7 Share of Documents current locations and Institutions*

Figures 4a and b highlighted the OA publication's present locations and institution collaborations ratio and its effect on average citations per paper. Of the OA publications, 83.42% of documents were presented in a single location. like journal websites and received 36.94% of total citations. Intitle “Application of genomics-assisted breeding for generation of climate resilient crops: progress and prospects” a (Gold OA) paper published in 2015 by Bidhan Chandra Krishi Viswavidyalaya (WB). Now, it is located in 15 places (repositories, websites, blogs). However, Figure 4b displayed that when the number of locations was changed, average citations also increased. It increased continuously in 2 locations 4762 papers received 10.41 average citations, 3 locations -2510-NP 16.82-ACP. The highest 61.70 ACP comes from 10+ locations from 10 papers. Figure 4a & b also presents institutions' collaboration ratio and citation advantage. A total of 141501 national and international Institutions were affiliated with this OA publishing. Of the scholarly outputs, 52.94% (n=34881) of records were published by only one institution and received overall citations of 26.31%. However, institutional collaborations also benefit from citations, which are presented in Figure 4b. Average citations increased from 4.19 in two institutions to 29.59 in 10+ institutions.

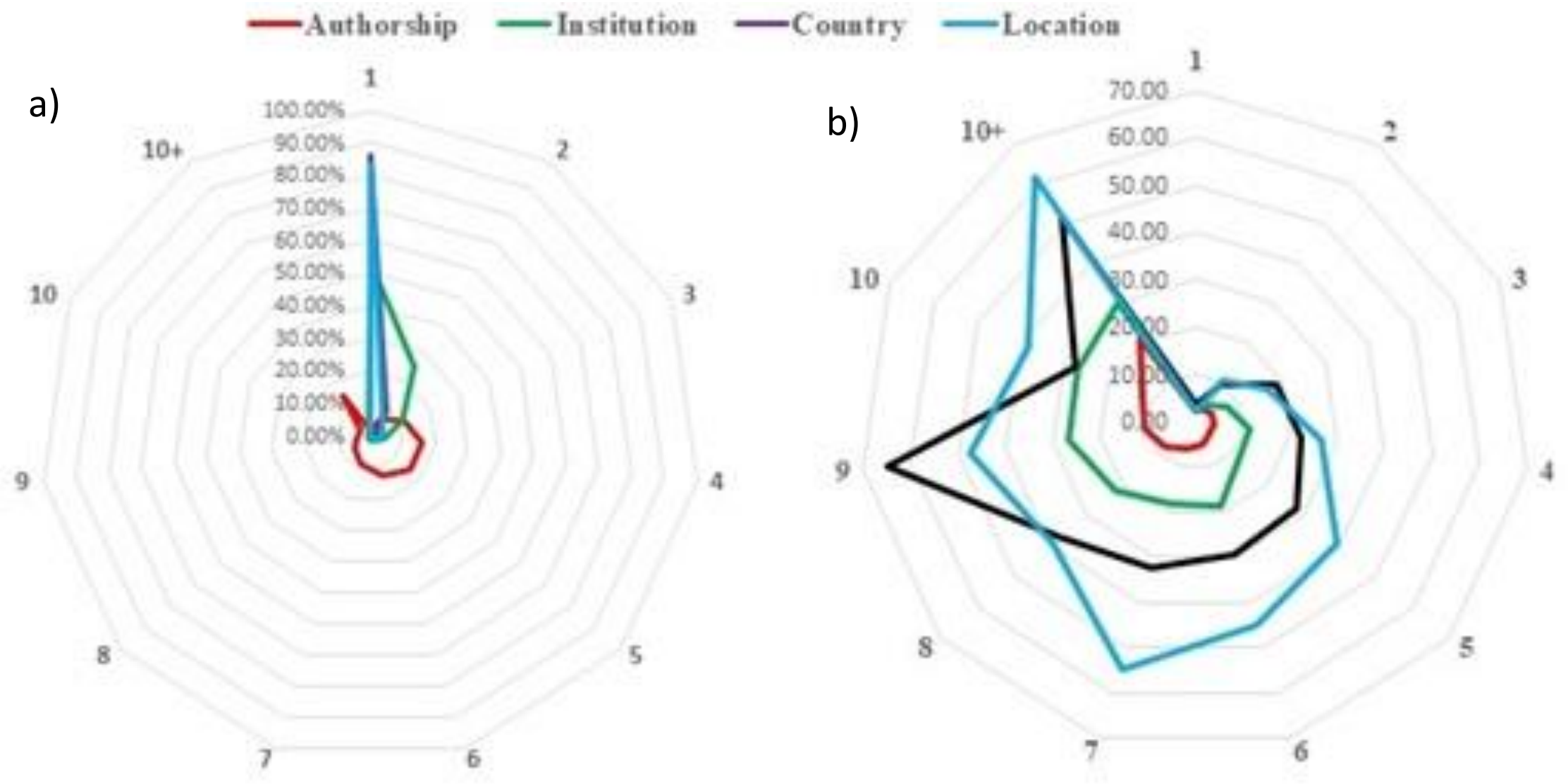

Figure 4: Pattern publications and average citations of authorship, Institutions, country and locations

### *3.8 OA colours growth at the State level*

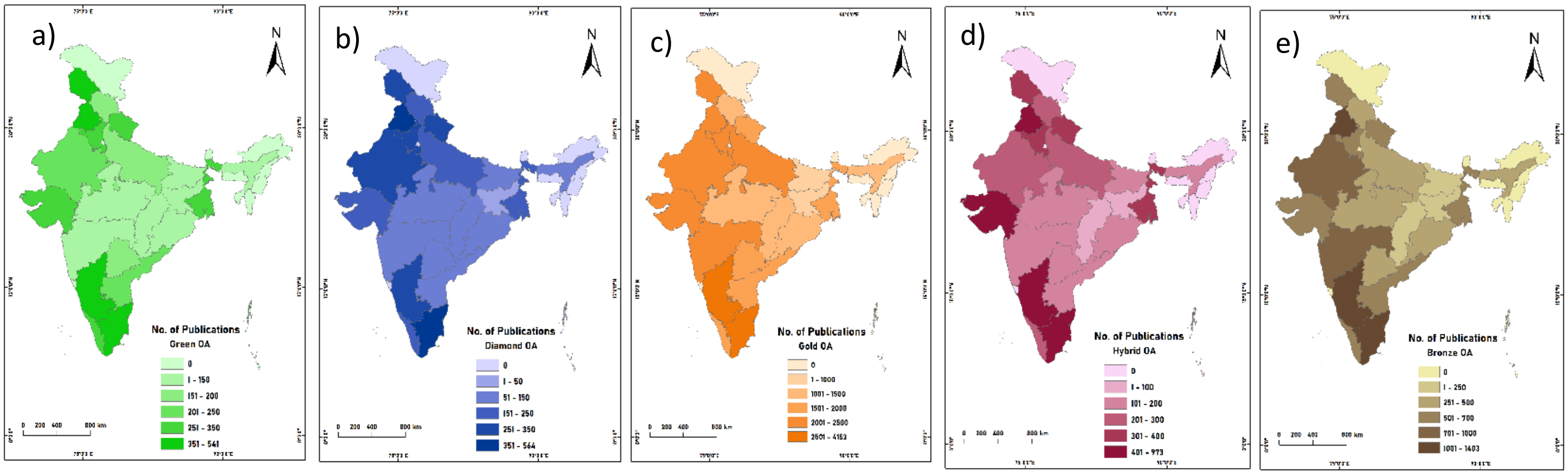

*Figure 5: Growth of OAs colours at the state level*

Figure 5 highlighted five OA-coloured growths at the state level. In the last ten years, the highest, 541 (11.17% of total Green OA) numbers of green OAs were distributed from Tamil Nadu (three state agriculture universities). This was followed by the second position achieved by Punjab (534) and the third position secured by Karnataka for publishing 380 green OA papers. Figure 5a points out the deep green parts in three states and Jammu and Kashmir, one union territorie also. Figure 5b bule map represented the Diamond OA publications state-wise overview. Punjab is the highest 564 (12.48%) Diamond OA productive state. Of the diamond OA publications, 554 papers were published in Tamil Nadu, it secured the 2nd position. The highest numbers of OAs came from Gold routes. Figure 5c orange Mape highlighted state-wise Gold OAs publishing performances. Tamil Nadu achieved the 1st position with publishing 4153 gold OA. Of the gold OA publications, 3947 were published from Karnataka, and it secured the 2nd position. Rajasthan, Uttar Pradesh, Maharashtra, Punjab, Gujarat, and Haryana, in five states, and Jammu & Kashmir union territories, were published in 2001-2500 on this scale. Map 5d maroon colour presents a hybrid OA publication state overview. Of the Hybide OA publications 15.98% (n= 973) of articles were shared from Tamil Nadu. Out of Tamil Nadu, Punjab- 695, Karnataka- 436, and Gujarat- 436 in three states secure places on 401-973 in this scale range. Olive colour mape 5e displayed bronze OA publishing trends. Tamil Nadu ranked in 1st position to publish 1403 bronze OA. Karnataka (1204) secured 2nd and Punjab (1040) 3rd positions, respectively, for publishing thousands of papers.

### *3.9 Funding status and impact of research*

Table 6 presents the top ten funing received universities' name and their scholarly impacts. In total, OA publishing 3.22% (n=2124) of OAs received funding support from two hundred national and international agencies. The top ten funding agencies are the Indian Council of Agricultural Research-311, Department of Biotechnology, Ministry of Science and Technology, India -252, Science and Engineering Research Board-151, Department of Science and Technology, Ministry of Science and Technology, India-114, King Saud University-85, Bill and Melinda Gates Foundation-76, University Grants Commission-71, National Natural Science Foundation of China-59, Council of Scientific and Industrial Research, India-47, National Research Foundation of Korea-45. Punjab Agricultural University has received the highest 214 (10.08% of 2124 papers) records of funding support among the 56 universities. Table 6 compares the impact of funding records with non-funding papers. Universities' Funding records received an average of 20 citations per paper. On the other side, non-funding papers received 4.51 average citations. University of Agricultural Sciences, Bangalore received the highest 25.08 of average citations papers. Of the 2124 funding publications shared with 635 journals. The top five journals are- PloS one-83, Frontiers in plant science-77, Scientific reports-77, 3 biotech-55, and Heliyon-36.

*Table 6: Top ten funds received university names with scholarly impact*

| Institutions name | *With Funding* | | | *Without Funding* | | |
|---|---|---|---|---|---|---|
| | *Papers %* | *TC* | *ACP* | *Papers* | *TC* | *ACP* |
| Punjab Agricultural University, Ludhiana | 214 (10.08) | 5076 | 23.72 | 3264 | 26051 | 7.98 |
| Sher-e-Kashmir University of Agricultural Science & Technology, Srinagar | 158 (7.44) | 3315 | 20.98 | 2345 | 18925 | 8.07 |
| Tamil Nadu Agricultural University, Coimbatore | 155 (7.30) | 3559 | 22.96 | 5554 | 18833 | 3.39 |
| G.B. Pant University of Agriculture & Technology, Pantnagar | 107 (5.04) | 2568 | 24.00 | 2831 | 28548 | 10.08 |
| University of Agricultural Sciences, Bangalore | 96 (4.52) | 2408 | 25.08 | 1487 | 7876 | 5.30 |
| Kerala Veterinary and Animal Sciences University, Kerala | 80 (3.77) | 1330 | 16.63 | 1386 | 5017 | 3.62 |
| Tamil Nadu Veterinary & Animal Sciences University, Chennai | 75 (3.53) | 1650 | 22.00 | 1469 | 6877 | 4.68 |
| Guru Angad Dev Veterinary and Animal Sciences University, Ludhiana | 71 (3.34) | 1486 | 20.93 | 1304 | 8121 | 6.23 |
| Chaudhary Charan Singh Haryana Agricultural University, Hisar | 68 (3.20) | 1414 | 20.79 | 2899 | 12525 | 4.32 |
| Assam Agricultural University, Jorhat | 68 (3.20) | 856 | 12.59 | 1880 | 5858 | 3.12 |
| Kerala University of Fisheries and Ocean Studies, Kochi | 58 (2.73) | 901 | 15.53 | 299 | 2313 | 7.74 |
| West Bengal University of Animal & Fishery Sciences, Kolkata | 58 (2.73) | 907 | 15.64 | 659 | 3982 | 6.04 |
| Others 44 universities | *916 (43.13)* | 17863 | 19.50 | 38388 | 142535 | 3.71 |
| *Subtotal* | *2124 (100%)* | 43333 | 20.40 | 63765 | 287461 | 4.51 |

Figure 6 points out the year-wise OA colours funding status and citation impacts. The Green OA publications 6.90% (334 out of 4843), hybrid OA 3.84%, Gold OA 3.34%, diamond OAs 1.91% and bronze OAs 1.63% of papers received Funding support. In the time period, 26 papers ( Gold-17, Hybrid- 3 and Bronze- 6) in 2014 to 337 papers (Green- 29, Diamond- 18, Gold – 226, Hybrid-46, and Bronze- 18) in 2023 were supported by funding agencies. The highest number of 472 papers, were received with finding support in the year 2022. The highest 81.27 average citations come from 11 hybrid OA papers published in 2015. It also highlights that hybrid, gold, bronze and diamond OA papers average citations higher than Green OA papers.

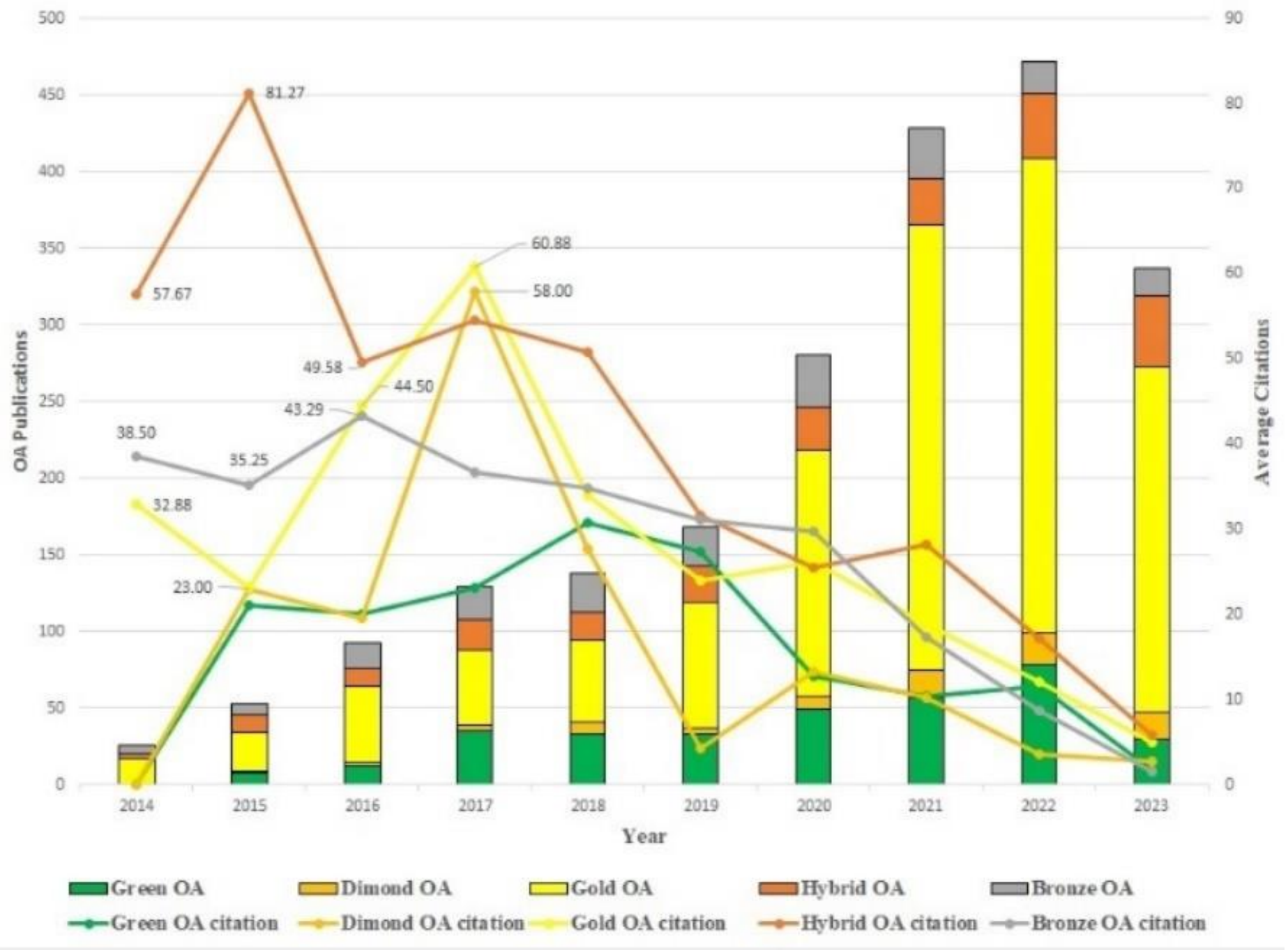

*Figure 6: Yearly growth of Funding Research and scholarly Impact*

*3.10 Use of Article processing charge and scholarly impact*

Table 7 presents a comparative overview of the top ten APC-used universities, as well as their time citations, ACP and used of USD currency. The Punjab Agricultural University, Ludhiana, has used 2191455 USD for the highest 869 papers. It received a 23.02 average citation. Among the top ten listed universities, the highest 34.57 ACP comes from 606 papers published by G.B. Pant University of Agriculture & Technology, Pantnagar. It also highlights that APC-used papers have average citations of 20.68, and without APC papers have average citations of only 2.33.

*Table 7: top ten APC-used university names with scholarly impact*

| Institutions name | With APC | | | | Without APC | | |
|---|---|---|---|---|---|---|---|
| | Papers % | TC | ACP | USD | Papers | TC | ACP |
| Punjab Agricultural University, Ludhiana | 869<br>9.00% | 20005 | 23.02 | 2191455 | 2609 | 11122 | 4.26 |
| Sher-e-Kashmir University of Agricultural Science & Technology, Srinagar | 633<br>6.55% | 15455 | 24.42 | 1448086 | 1870 | 6785 | 3.63 |
| G.B. Pant University of Agriculture & Technology, Pantnagar | 606<br>6.27% | 20949 | 34.57 | 1301773 | 2332 | 10167 | 4.36 |
| Tamil Nadu Agricultural University, Coimbatore | 590<br>6.11% | 11974 | 20.29 | 1319423 | 5119 | 10418 | 2.04 |
| Kerala Veterinary and Animal Sciences University, Kerala | 444<br>4.60% | 3923 | 8.84 | 511145 | 1022 | 2424 | 2.37 |
| Chaudhary Charan Singh Haryana Agricultural University, Hisar | 389<br>4.03% | 7707 | 19.81 | 887812 | 2578 | 6232 | 2.42 |
| Guru Angad Dev Veterinary and Animal Sciences University, Ludhiana | 366 3.79% | 6865 | 18.76 | 760134 | 1009 | 2742 | 2.72 |
| Orissa University of Agricultural & Technology, Bhubaneswar | 285 2.95% | 4857 | 17.04 | 585450 | 1659 | 3928 | 2.37 |
| Tamil Nadu Veterinary & Animal Sciences University, Chennai | 282 2.92% | 5782 | 20.50 | 589095 | 1262 | 2745 | 2.18 |
| University of Agricultural Sciences, Bangalore | 261 2.70% | 6620 | 25.36 | 653483 | 1322 | 3664 | 2.77 |
| Bidhan Chandra Krishi Viswa Vidhyalaya, Mohanpur | 261 2.70% | 6507 | 24.93 | 603822 | 1311 | 3779 | 2.88 |
| Others 45 universities | 4672 48.37% | 89098 | 19.07 | 9650642 | 34138 | 67046 | 1.96 |
| Subtotal | 9658 100% | 199742 | 20.68 | 20502320 | 56231 | 131052 | 2.33 |

Figure 7 points out the tendency of APC to use in the last ten years. A total of 14.66% of OA papers (9658 out of 65889) were paid the APC charge. Diamond OA published without APC charge. green OAs are 25.83%, Gold OAs 18.08%, Hybrid OAs -9.45% and bronze OA 7.90% papers paid APC according to OpenAlex. In 2014, 531 papers (green- 122, gold- 314, hybrid- 41 and bronze- 54) paid APC charge. It was increased continuously. it over 1500 papers in 2021, 2022 and 2023. It was also highlighted that in the first nine years, APC paid papers with an Average citation of 10 or above.

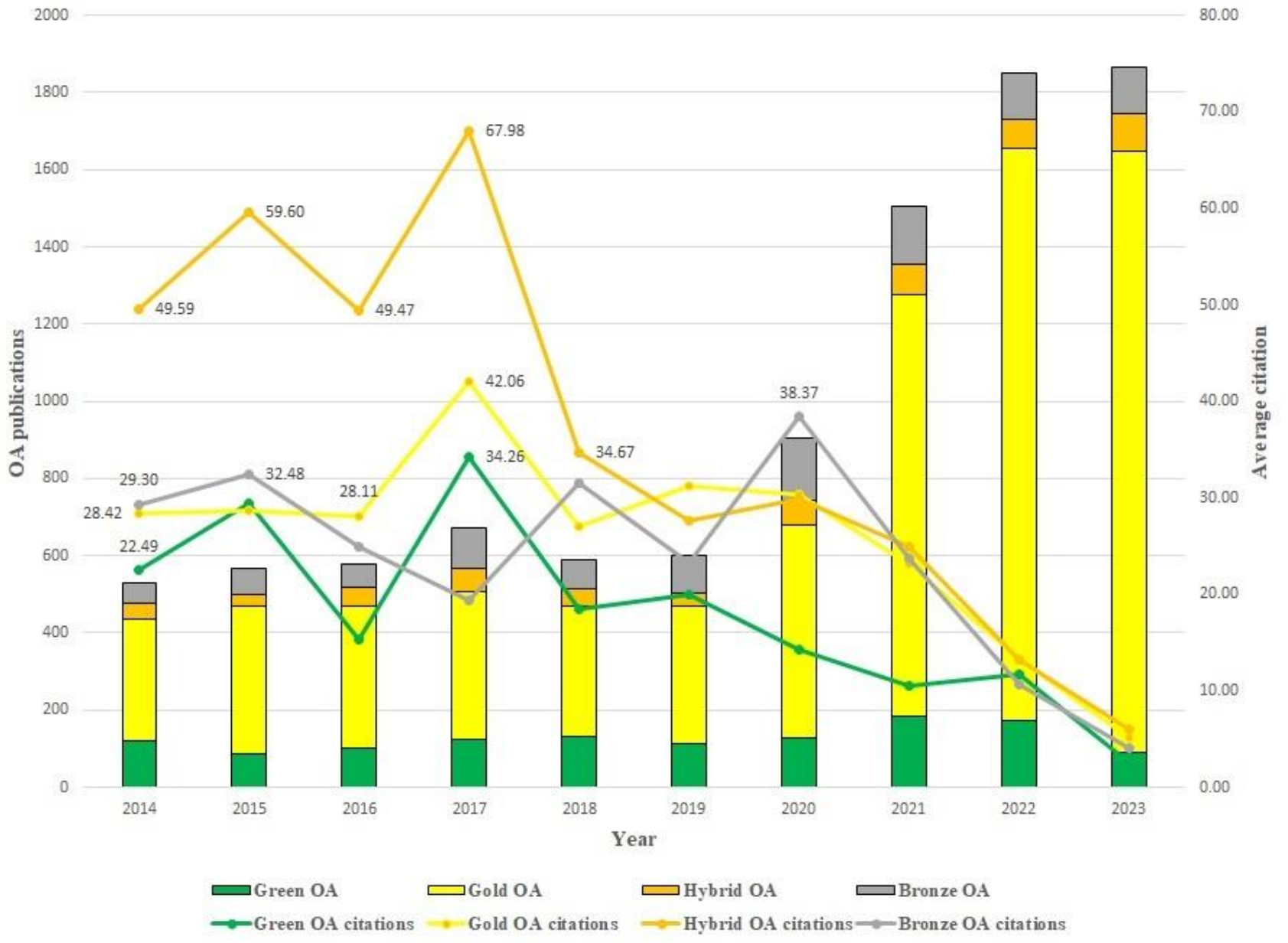

*Figure 7: Yearly used APC and scholarly Impact*

*3.11 Most frequency Keywords*

Figure 8 displays the top twenty keywords that occur above five hundred. The top useable keyword is Crop Productivity (Occurrence 3023). This is followed by soil fertility, which is the second most useable keyword (occurrence- 2941), and the third one is Grain Yield (Occurrence-2223); almost all twenty keywords are related to agriculture.

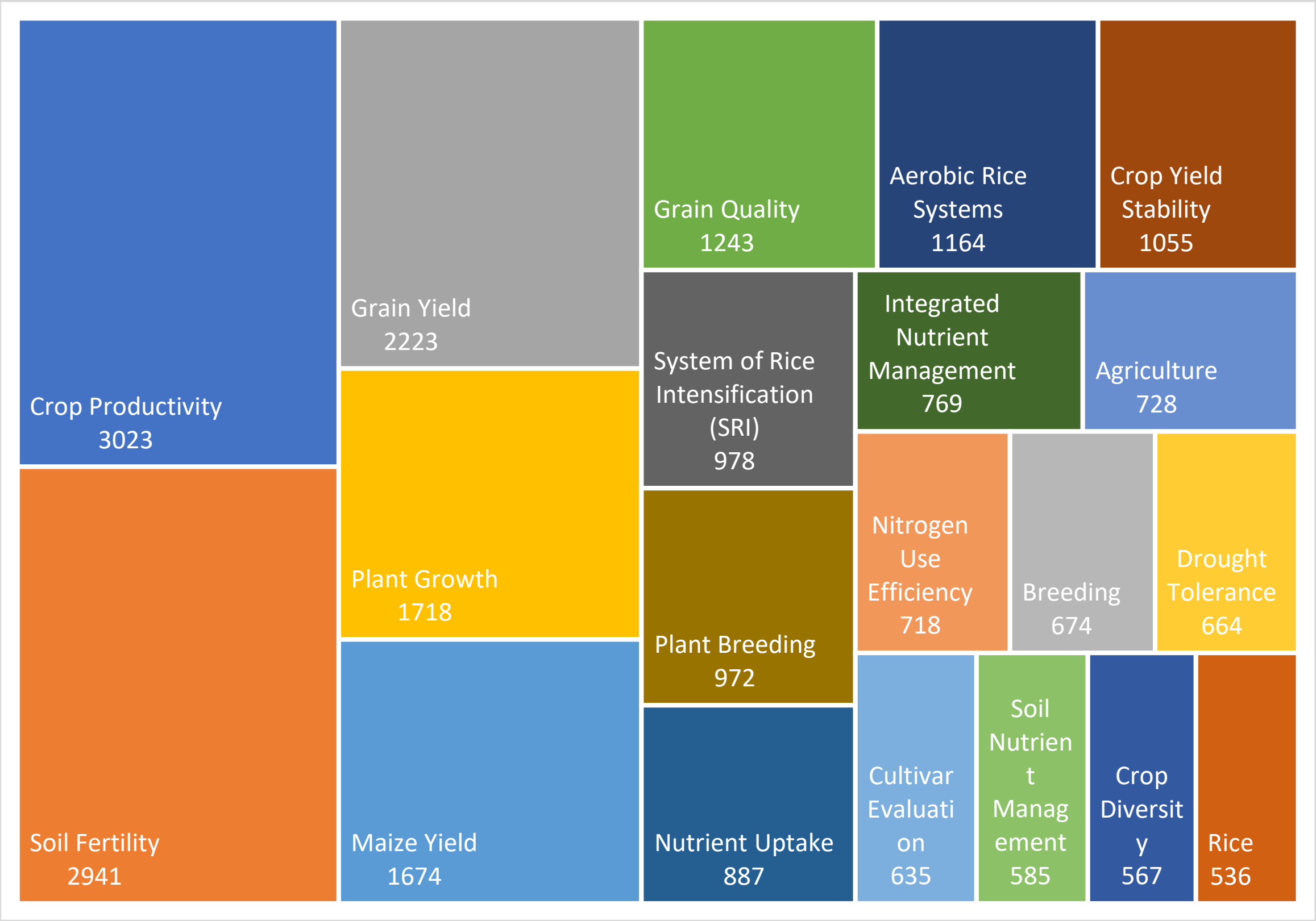

*Figure 8: Top useable keywords in the agriculture domain*

*3.12 Domain and topic*

Figure 9a & b displayed domains related to OA publications and average citations. In OA publications, 74.09% (48817 out of 65889) of papers are related to Life Sciences, 11.77% to Physical Sciences, 10.92% are Health Sciences, 3.00% belong to Social Sciences, and 0.22% are others. Five OA colours also like the same trends. However, when this study observed the citation trend of OA colours, it found that the highest average citations came from health science (4-8 ACP) and physical science (5-10 ACP) per paper. Interestingly, Green and Hybrid OA papers received average citations of 6.01 and 5.00, respectively. Only the diamond OA papers receive 5.62 average citations in the social science group.

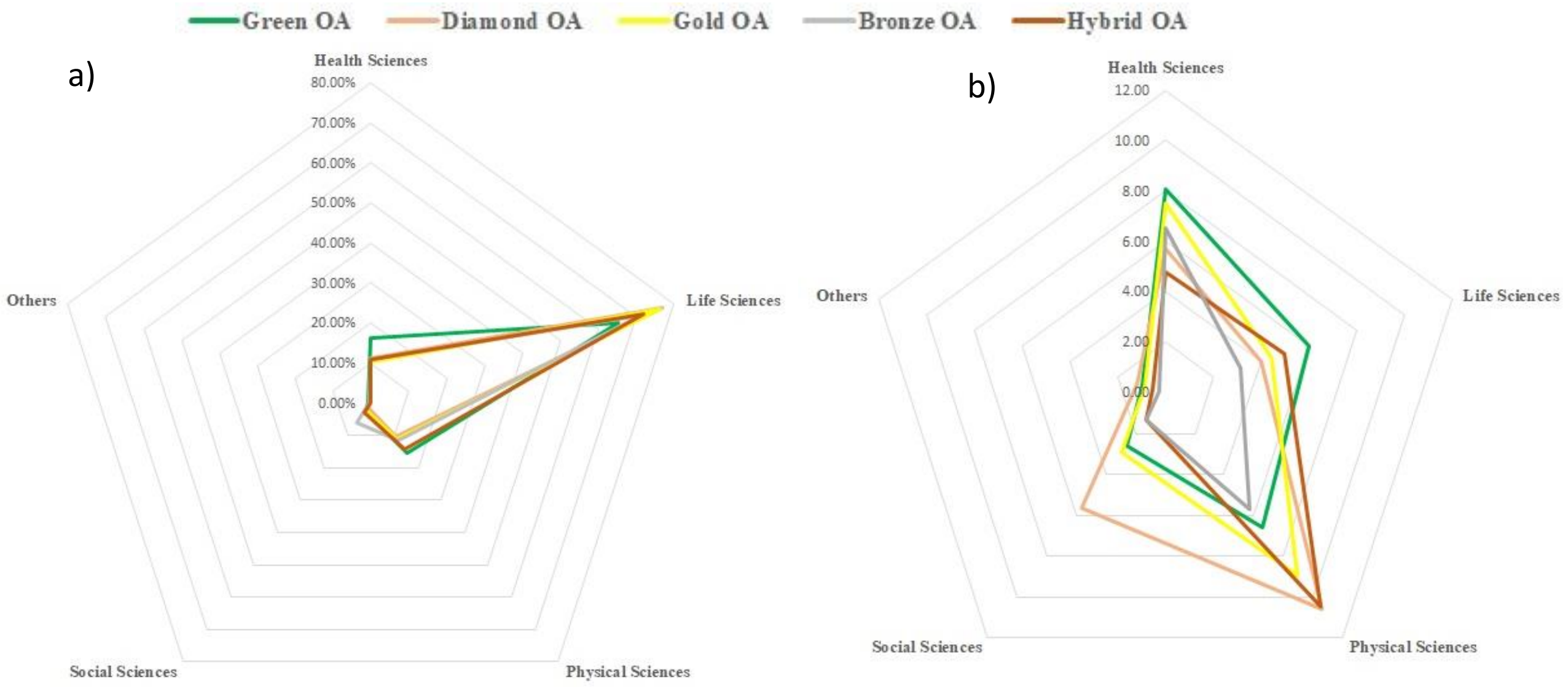

*Figure 9: OA papers coverage with domain and Average citation*

Table 8 tabulated the twenty-one topics' names, total publications, time citations and average citations per paper. A total of 2171 topics were found in this research. Among them, five hundred enlisted papers related to topics are displayed here. The highest number of papers (3881) related to Agricultural Development and Policy in South Asia. This is followed by 2556 papers connected to Rice Water Management and Productivity Enhancement, and the third one is Management of Soil Fertility and Crop Productivity. The Rice Water Management and Productivity Enhancement have received the highest of 9663 citations. The Mechanisms of Plant Immune Response, this topic 560 papers have received the highest 12.44 average citation per paper among this group.

*Table 8: Trending topic in the Agriculture Domain*

| *SL* | *Topics name* | *TP* | *TC* | *ACP* |
|---|---|---|---|---|
| 1 | Agricultural Development and Policy in South Asia | 3881 | 3638 | 0.94 |
| 2 | Rice Water Management and Productivity Enhancement | 2556 | 9663 | 3.78 |
| 3 | Management of Soil Fertility and Crop Productivity | 1996 | 3063 | 1.53 |
| 4 | Genetics and Breeding of Cowpea | 1488 | 2933 | 1.97 |
| 5 | Factors Affecting Maize Yield and Lodging Resistance | 1469 | 2847 | 1.94 |
| 6 | Cultivar Evaluation and Mega-Environment Investigation | 1338 | 4564 | 3.41 |
| 7 | Genetic Diversity and Breeding of Okra | 1224 | 1752 | 1.43 |
| 8 | Physiology and Management of Fruit Trees | 1077 | 1918 | 1.78 |
| 9 | Livestock Farming and Rural Development Practices | 1013 | 1028 | 1.01 |
| 10 | Genomics and Breeding of Legume Crops | 833 | 5265 | 6.32 |
| 11 | Intercropping in Agricultural Systems | 826 | 2283 | 2.76 |
| 12 | Diversity and Evolution of Fungal Pathogens | 754 | 1823 | 2.42 |
| 13 | Deficit Irrigation for Agricultural Water Management | 678 | 1321 | 1.95 |
| 14 | Biofortification of Staple Crops for Human Nutrition | 676 | 6138 | 9.08 |
| 15 | Mechanism of Plant Growth Regulation | 665 | 832 | 1.25 |
| 16 | Mechanisms of Plant Immune Response | 560 | 6967 | 12.44 |
| 17 | Elicitor Signal Transduction for Metabolite Production | 554 | 2153 | 3.89 |
| 18 | Health Benefits of Wheatgrass Consumption | 547 | 1147 | 2.10 |
| 19 | Viral RNA Silencing and Plant Immunity | 544 | 3827 | 7.03 |
| 20 | Reproductive Health in Dairy Cattle | 534 | 998 | 1.87 |
| 21 | Genomic Selection in Plant and Animal Breeding | 532 | 1773 | 3.33 |
| | Others 2150 topics | 42144 | 264861 | |
| | Total topics 2171 | 65889 | 330794 | |

*4 Findings and Discussion*

This comprehensive analysis provides valuable insights into the OA research landscape of state agriculture universities in India. The results reveal that OA publications deep insight, including publication trends, most productive state, prolific Institution, top authors, leading journals, funding supports and use of APC.

The increasing trends of OA publishing in the last ten years give us a positive message. Authors are concerned about OA publishing routes. Highly active institutions, authors and states provide a comprehensive overview of research on sustainability in agriculture. Tamil Nadu Agricultural University in Coimbatore is a leading institution in publishing research through open-access routes. This achievement was made possible through the cooperative efforts of authors and institutions. The rankings of three authors—M. Raveendran (158 publications), V. Geethalakshmi (92 publications), and N. Senthil (90 publications)—serve as evidence of this collaborative success. Followed by Punjab Agricultural University, Chaudhary Charan Singh Haryana Agricultural University, Hisar, G.B. Pant University of Agriculture & Technology and Sher-e-Kashmir University of Agricultural Science & Technology, Srinagar are shared notable OA contributions. Similarly, the collaborative efforts of Tamil Nadu Agricultural University in Coimbatore (5,709 publications), Tamil Nadu Veterinary and Animal Sciences University in Chennai (1,544 publications), and Tamil Nadu Dr. J. Jayalalithaa Fisheries University in Nagapattinam (371 publications) have positioned Tamil Nadu as the most productive state in India. Based on the evidence of OA percentage, the analysis classified five OAs: Green OA (7.35% average citation 6.32), Diamond OA (6.74% -4.93), Gold OA (57.27%-5.21), Hybrid OA (9.24%-5.64) and Bronze OA (19.41%-3.70). Indian State Agriculture Universities' OA publications percentage is high compared to Non-Profit organizations with (57.74%) (Hazarika et al., 2024), Central universities with 28.43% (Roy & Mukhopadhyay, 2022b), state universities with 24.57% (Roy & Mukhopadhyay, 2022c).

Still, the research finds that there is a great difference between domestic journals and foreign journal publications' ratio and average citations (Garg et al., 2006). International Journal of Current Microbiology and Applied Sciences, International Journal of Chemical Studies and International Journal of Environment and Climate Change are three Indian journals that play an important role in OA publications. The authors collaborate with U.S. institutions and publish their papers in American journals (Nidhisha & Sarangapani, 2021; Tripathi et al., 2013; Tripathi & Garg, 2014).

A closer examination of the countries, authors, institutions, and locations associated with shared papers' citation advantages reveals some intriguing findings. The number of authors, institutions, countries, and locations contributes to the sustainability of knowledge and increases readers' interest, leading to more citations (Hazarika et al., 2024).

The analysis of funding status reveals the lack of support in agriculture research. Only 3.22% (n=2124) of OA papers were published with funding support. It was observed that the Government of India archives 7th position in support of research development funds and The Department of Science and Technology-Funding for Improvement of Science and Technology (DST-FIST)-(4%) supported funding in Agriculture Research (Srinivasaiah et al., 2021).

Lastly, trending topics highlight the recent developments in agricultural research in India. Since global warming directly influences climate change worldwide, scholars are focusing on agricultural development policies, rice water management and productivity, as well as the management of soil fertility and crop productivity.

*5 Conclusion*

This experiment offers a comprehensive overview of the open-access publishing trends, international collaborations, authorship patterns, and funding status of Indian State Agricultural Universities. It highlights opportunities for institutions, state governments, and funding agencies to prioritise open-access publishing to promote sustainable agricultural research. Additionally, researchers are encouraged to develop new indicators to measure research transparency.